# Mapping Moral Valence of Tweets Following the Killing of George Floyd

J. Hunter Priniski,[1] Negar Mokhberian,[2] Bahareh Harandizadeh,[3] Fred Morstatter,[4]
Kristina Lerman,[5] Hongjing Lu,[6] P. Jeffrey Brantingham[7]

Department of Psychology[1,6], Department of Statistics[6], Department of Anthropology[8], University of California, Los Angeles
Department of Computer Science[2,3], Information Sciences Institute[4,5], University of Southern California
{priniski, hongjing, branting}@ucla.edu, {nmokhber, harandiz, morstatt, lerman}@usc.edu

**Abstract**

The viral video documenting the killing of George Floyd by Minneapolis police officer Derek Chauvin inspired nation-wide protests that brought national attention to widespread racial injustice and biased policing practices towards black communities in the United States. The use of social media by the Black Lives Matter movement was a primary route for activists to promote the cause and organize over 1,400 protests across the country. Recent research argues that moral discussions on social media are a catalyst for social change. This study sought to shed light on the moral dynamics shaping Black Lives Matter Twitter discussions by analyzing over 40,000 Tweets geo-located to Los Angeles. The goal of this study is to (1) develop computational techniques for mapping the structure of moral discourse on Twitter and (2) understand the connections between social media activism and protest.

## Introduction

George Floyd's murder on May 25th, 2020 inspired national protests which brought unprecedented attention to police brutality and racial injustice. As an estimated 26 million Americans participated in the protests over the first month (Hamel, Kearney, Kirzinger, Lopes, Munana, & Brodie, 2020), it is considered the largest social movement in American history (Buchanan, Bul, & Patel, 2020). The movement was catalyzed by Black Lives Matter activists, who through a combination of public protest and social media activism, dramatically increased White communities' awareness of issues of racial injustice and disparities in policing (Buchanan, et al., 2020).

Discussion about police brutality echoed on social media as well. Tweets with the #BlackLivesMatter hashtag inspired social media discussions about police brutality and racial injustice since its inception in 2013 (Anderson, Barthel, Perrin, & Vogels, 2020). After George Floyd's death, the #BlackLivesMatter movement brought scenes of police brutality to millions of people's social media feeds and dramatically increased awareness of the movement. On May 28th, 2020, three days after George Floyd's death, the hashtag had been used a total of 8.8 million times, the largest spike in its history (Anderson, et al., 2020). In the age of the internet, social media can promote social causes to vast audiences (LeFebvre & Armstrong, 2018). It is therefore important to understand the nature of these discussions.

Previous research has articulated the relationship between social media discussions and protests focused on unjust policing practices. For instance, the amount of moral rhetoric in Tweets about the 2015 Baltimore police killing of Freddie Gray predicted the number of violent demonstrations and subsequent arrests (Mooijman, Hoover, Lin, Ji, & Dehgani, 2018). More generally, social media users expressing moral outrage plays a key role in social movements in the digital age, as doing so increases group bonds (in-group connectivity) and inspires a tight collective of people to work towards a social cause (Brady, Crockett, & Van Bavel, 2020). As a consequence, we would expect that distinct moral considerations can be measured in Black Lives Matter Twitter discussions in addition to language that expresses positive attitudes towards ingroup membership. This is because moral outrage can create common knowledge around a social issue, form cohesive social groups, and inspire collective activism (Spring, Cameron, & Cikara, 2018).

In this work, we analyze the moral evaluations Twitter users made in response to George Floyd's murder through Moral Foundations Theory (MFT), a framework for explaining variation in people's moral reasoning (Graham, et al., 2013). The framework decomposes the types of moral evaluations people make into five foundations: Authority/Subversion, Care/Harm, Fairness/Cheating, Loyalty/Betrayal, Sanctity/Purity (see Table 1). Moral Foundations Theory explains the presence of and variation in moral sentiments as the product of innate cognitive modules that are shaped by local cultural contexts and which can be reduced to one of the five foundations (Haidt & Joseph, 2004). Consequently, moral disagreements about politicized issues can emerge because Liberals and Conservatives rely on differing sets of moral foundations (Haidt & Nosek, 2009).

The emphasis of moral foundations is most commonly inferred from written text (speech acts) by flagging combinations of words that have validated connections to each foundation (Graham, Haidt, & Nosek, 2009; Frimer, Skitka, & Motyl, 2017; Kennedy, et al., 2021). For instance, terms such as "justice" and "trustworthiness" are related to the Fairness foundation and "patriotism" is related to Loyalty. Recent behavioral research has focused on developing extended vocabulary sets with human ratings for mapping large sets of terms onto various moral foundations (Hopp, Jisher, Cornell, Husky, & Weber, 2021). Such dictionaries serve as the backbone for computational methods for extracting moral sentiment from natural language text (see Kennedy, Ashokkumar, Boyd, & Dehghani, 2021 for a review).

In this paper, we apply a FrameAxis approach to represent the moral sentiment in Tweets (Kwak, An, Jing, & Ahn, 2020). This approach calculates the cosine similarities between the set of moral words in the document by leveraging the moral foundation ratings for terms in the extended Moral Foundations Dictionary (Hopps, et al., 2021). This method results in a moral "framing" for each document in terms of its intensity (i.e., degree to which that foundation is expressed) and bias (virtue or vice) for each moral foundation. Documents (here, Tweets) are mapped into a *moral embedding space*, where Tweets that are close to one another in this space are assumed to express similar moral sentiment. Our approach is well-suited for representing the moral valence of short texts because it leverages semantic distances between the words, which more traditional unsupervised methods do not do (e.g., Hopp et al., 2021). With this method, we aim to answer three general questions in this paper:

1. What effect did the killing of George Floyd have on catalyzing the BLM movement on Twitter?
2. What moral considerations in discourse inspire civic activism?
3. Do moral structures emerge in the discussion channels?

## Methods

**Data Collection**
We collected 36,282 Tweets geo-located to the greater Los Angeles area (bounding box: 32.75, -118.95 and 34.82, -117.646374). The dataset includes geo-coded Tweets, where latitude and longitude associated with a post is known, and Tweets where a place (e.g., city) is associated with all of a user's posts. The collection period spanned February 24, 2020, to August 24, 2020.

We then searched for Tweets matching at least one of the following search terms: #blacklivesmatter (resulting Tweet count = 15,499), blm (14,432), #blm (5,807), "black lives matter" (4,766), #defundthepolice (1,769), "all lives matter" (1,566), "defund the police" (1,039), #alllivesmatter (549), #bluelivesmatter (305).

## Mapping Tweets to Moral Embedding Space

The first step towards representing the moral sentiment of the Tweets in our dataset is to map Tweets into a moral embedding space, where the position of each Tweet in the space represents the document's moral sentiment with respect to the five moral foundations. We apply the FrameAxis approach with moral terms from the extended Moral Foundations Dictionary.

**FrameAxis Modeling of Moral Sentiment**
The FrameAxis method defines *semantic axes* in the latent space of word embeddings and then calculates the relevance of any given text to those axes. The axes are built by employing the SemAxis approach (An, Kwak, & Ahn, 2018), where opposing sets of words are used to build a meaningful semantic axis. Here, the semantic axes correspond to the five moral foundations, and the sets of terms associated to each foundation's vice and virtue domain are leveraged to build each *moral foundation axis* (Reiter-Haas et al. 2021, Mokhberian et al. 2020). For instance, the Care/Harm axis is defined with virtue words such as "care", "help", "provide", etc. at one pole, and vice words such as "attack", "violence", "kill", etc. at the other pole.

We use the set of human rated terms in the extended Moral Foundations Dictionary to form the moral foundation axes. Each word in the dictionary is assigned five probability values representing the probability the term is relevant to each of the five moral foundations, the term is labeled as the foundation with the highest probability value. Next, a sentiment analysis tool is used to assign the terms to the vice (negative) or virtue (positive) dimension. We then uncover the moral foundation axes by calculating the difference vector between the centroids of that foundation's cluster of vice and virtuous terms.

For mapping a Tweet to the moral embedding space, FrameAxis calculates the *bias* and *intensity* of each Tweet toward each of the moral foundation axes. Bias represents the valence of the Tweet in a moral foundation category. Positive bias values imply the Tweet is more virtuous, while negative values imply alignment with the vice dimension. A Tweet's intensity represents how relevant the Tweet is to each moral foundation. If the Tweet contains many words with high cosine similarity with a moral foundation axis, then the Tweet's relevance to that foundation will be high.

Using this approach, we find a vector representation of each Tweet's text such that each element of the vector represents the degree to which the Tweet is related to each of the five moral foundations in the vice (negative) or virtue (positive) domain. Because each moral foundation has dictionary terms associated with them in the virtue and vice domain, this approach represents Tweet texts as 10-dimensional vectors. Descriptions of the moral foundations can be found in Table 1.

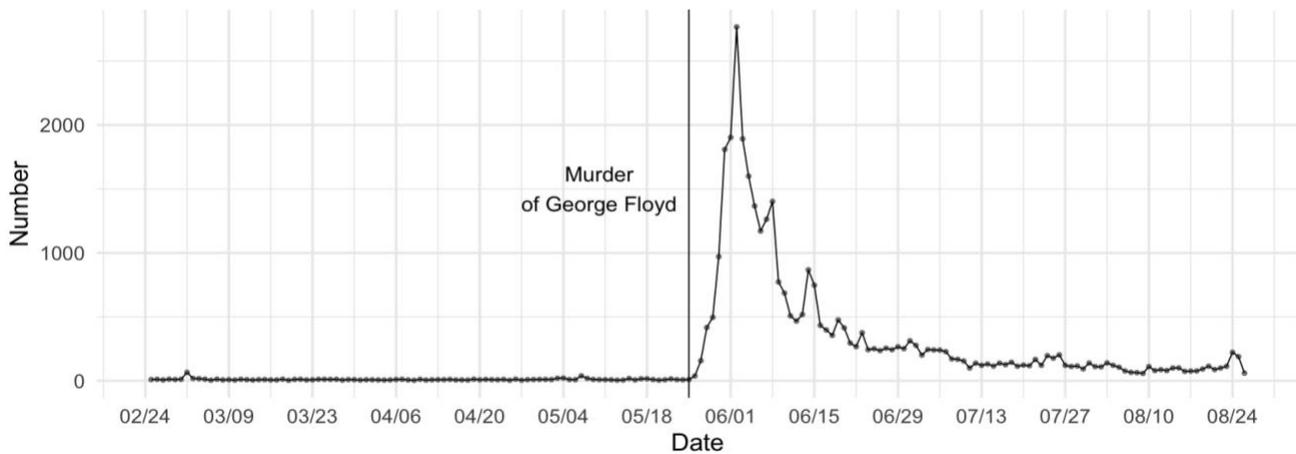

Figure 1. Daily counts of Black Lives Matter Tweets before and after the killing of George Floyd.

| Foundation (virtue/vice) | Description (Virtuous Direction) |
|---|---|
| Authority/ Subversion | Desire/need for beneficial relationships with hierarchies in society. |
| Care/ Harm | Compassion towards victims and the vulnerable, anger towards those perpetrating injustice and harm. |
| Fairness/ Cheating | Desire for cooperation and gratitude for just and trustworthy systems and people. |
| Loyalty, Ingroup/ Betrayal | Desire for cohesive groups. Instantiates group pride and anger at traitors. |
| Sanctity, Purity/ Degradation | Relevant virtues consist of being temperamental and pious and clean. |

Table 1. Description of moral foundations under Moral Foundations Theory. Table adapted from Haidt (2012) and Graham, et al. (2013).

### Examining Structure of Moral Embeddings

After mapping the text of each Tweet to the moral embedding space, we can then perform statistics on these embeddings to examine what types of moral considerations are most prominent. First, we calculate the average activation (i.e., intensity value) for each foundation value, which allows us to assess the most prominent moral evaluations in the corpus and to infer which considerations are most prominent in collective organizing on Twitter.

While taking the mean activations for each foundation can be informative about the general trends in the data, Twitter data is rich and there is likely to be more complex structure to the data topology. To this end, we used the $k$-means clustering algorithm in the Python library scikit-learn to uncover clusters of documents in the moral embedding space (Pedregosa, et al., 2011).

We hypothesize $k$-means clustering will afford a finer understanding of how data is mapped in the space and reveal the presence of distinct sets of moral considerations.

### Results

We break the study's Results into the following subsections: (1) **Extracting Moral Tweets** (assessment of how the killing of George Floyd catalyzed moral discourse on Twitter), (2) **Assessing Moral Foundation Activation** (which moral foundations are most at play?) and (3) **Exploring Data Structure in Embedding Space** (examine if there is rich structure to how documents are distributed in this space). Taken together, these analyses will provide an initial glimpse at how moralized Twitter discussions shaped Black Lives Matter activism after George Floyd's death.

### Extracting Moral Tweets

Tweets that didn't contain any of the terms in the extended Moral Foundations Dictionary were excluded from analysis. Tweets that contain at least one term are coined *moral Tweets*. Table 2 reports the number of moral Tweets with unique text strings in the dataset. As seen in Figure 1 (above), close to all of these moral Tweets were posted after the killing of George Floyd. In line with a large body of previous findings, this result suggests the event catalyzed #BlackLivesMatter discussions to all-time highs.

| Hashtag | Number of Unique Tweets |
| --- | --- |
| Black Lives Matter | 26,827 |
| All/Blue Lives Matter | 1,578 |
| Defund the Police | 1,818 |
| Total | 30,223 |

Table 2. Unique Tweets in corpus determined by two embedding algorithms. Groups in the first column are determined by the content of the hashtag contained in the Tweet.

**Examining Moral Foundation Activation**

The main focus of this paper is to examine the moral discourse shaping the #BlackLivesMatter movement on Twitter. We sought to do this by measuring the mean levels of moral foundation activation for each Tweet in our dataset. As shown in Figure 2, the most active moral foundations are in the Authority, Care/Harm, and Fairness- vice domain. This is contrasted with high virtuous activations for Loyalty (or one's ingroup members) and Sanctity. While the high vice activations for Authority, Care/Harm and Fairness may be expected from the Black Lives Matter movement, which is by definition critique the authority of the police force and how they care for communities of color, the virtuous activations for Loyalty and Sanctity are more striking.

Inspection of the important terms associated with each foundation can provide richer context for what the moral discussions are about. Table 3 reports some terms closest to each foundation. Terms associated with the Authority (i.e., "violence" and "protest") and Care/Harm (i.e., "attack," "racist,", "police" and "killing") foundations reveal that discussants are likely expressing critiques of the police's use of violent force against protestors and communities of color. Terms associated with the Fairness foundation reveal that discussions of police behavior centers on racial discrimination, injustice, and bias in policing ("discrimination", "bias," "racism," "unfair," and "injustice").

We also see high virtuous activation for the Loyalty and Sanctity domains. This suggests expression of positive attitudes towards ingroup membership—as predicted by models of how moralized discussions catalyze activism— where positive ingroup attitudes emerge as a shared moral outrage takes hold surrounding an issue (Brady, et al., 2020). Indeed, associated terms suggest people are valuing "loyalty," "community," and "togetherness." The positive valence of Sanctity tells a similar story, where the terms reveal the role that religious group identity (i.e., 'Catholic' and 'Christian') and collective practices play for overcoming misjustice (i.e., 'prayer').

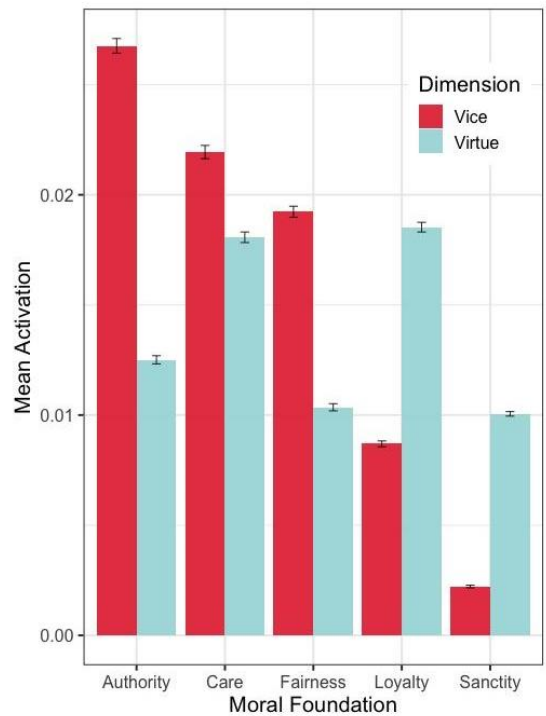

Figure 2. Mean activation of moral foundation axes for all Tweets in corpus. Errors bars represent 95% confidence intervals.

| Moral Foundation | Important Vocabulary |
| --- | --- |
| Authority | protest, demonstrators, violent, riot, arrested |
| Care/Harm | attack, killing, racist, blaming, riots, protests |
| Fairness | discrimination, bias, racism, unfair, injustice, hate |
| Loyalty/Ingroup | founding, together, community, unite, honor, proud |
| Sanctity/Purity | church, christian, prayer, catholic |

Table 3. Selected important vocabulary terms for deriving moral foundation activations by FrameAxis approach.

**Exploring Data Structure in Embedding Space**

While analysis of prominent moral activations and their associated terms provides an initial glimpse into the moral discussions taking place, a closer examination of how data is distributed throughout the embedding space is needed to gain a clearer understanding of the various types of present moral evaluations. To this end, we use clustering analyses to explore whether natural groups of documents are present in the embedding space.

We performed *k*-means clustering of the Tweets to reveal such structures. Model selection was guided by (1) choosing a value of *k* with a "good" silhouette coefficient (see below) and (2) which resulted in clusters with clearly differentiable distribution of moral foundation activations. The silhouette coefficient for a clustering model stems from the average intra-cluster distance and the nearest non-membership cluster distance for each sample in the dataset (Rousseeuw, 1987). Consequently, the silhouette coefficient provides a good metric for how tightly connected and separated the resulting clusters are. We chose a *k* value that resulted in clusters with roughly equal size and a high average silhouette coefficient. We further explored if the resulting clusters had clearly differentiable distributions of moral foundation activations. The preferred model based on these criteria has *k* = 4 clusters.

Figure 3 shows a t-SNE visualization (Van der Maaten and Hinton 2008) of Tweets in the embedding space, colored by their clusters under the *k*-means model. t-SNE approximates the spatial relationships that exist in the high dimensional moral embedding space as it maps points in two-dimensional space. We see that three large clusters emerge (Cluster 1, 2, and 3) in addition to one smaller cluster (Cluster 4).

Figure 4 communicates the average foundation activation of Tweets in each cluster, which allows us assess if each cluster maps differentially onto the moral foundation axes. These cluster-specific activations reveal two key findings: First, two separate clusters express more virtuous sentiment (i.e., Clusters 1 and 3) while Clusters 2 and 4 express more centered around the moral vices. Furthermore, comparing the activation intensities for Clusters 2 and 4 reveals that the smaller set of tweets express the highest intensities. Comparing the foundation levels across different clusters enables exploration of the types of moral considerations present in the corpus with greater specificity than simply looking at activations averaged across the corpus.

To begin exploring the content of each cluster, we compare the most frequently unique terms in each cluster. Table 5 reports a selected set of vocabulary words that frequently and uniquely appear in each cluster. We applied no specific cutoff rule for presenting these terms as this step is largely exploratory.

Terms like "blacklivesmatter", "police", and "blm" were common to each cluster. However, terms orienting around geographic locations (i.e., Hollywood, Los Angeles California) are common in Clusters 1 and 2. This suggests that these clusters contain Tweets focused on community organizing, while in Cluster 3 we see more general supportive terms for the movement (i.e., "good", "change," and "support"). The most morally active cluster, Cluster 4, seems to be Tweets focused on promoting the cause on social media.

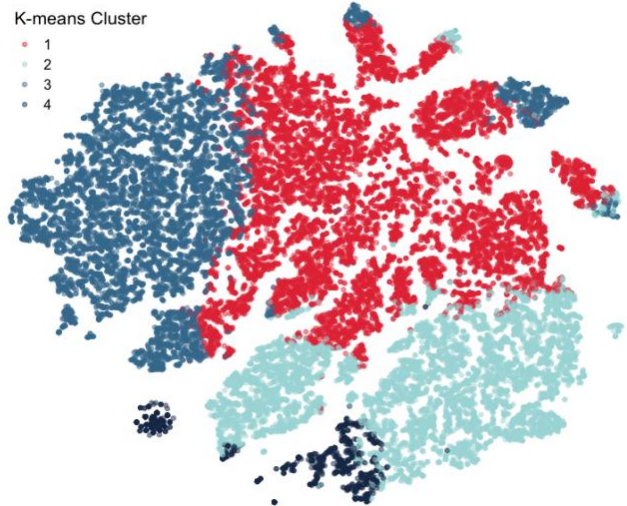

Figure 3. t-SNE visualization of Tweets in moral foundation embedding space. individual points represent individual Tweets, and Tweets are collared by their *k*-means cluster label.

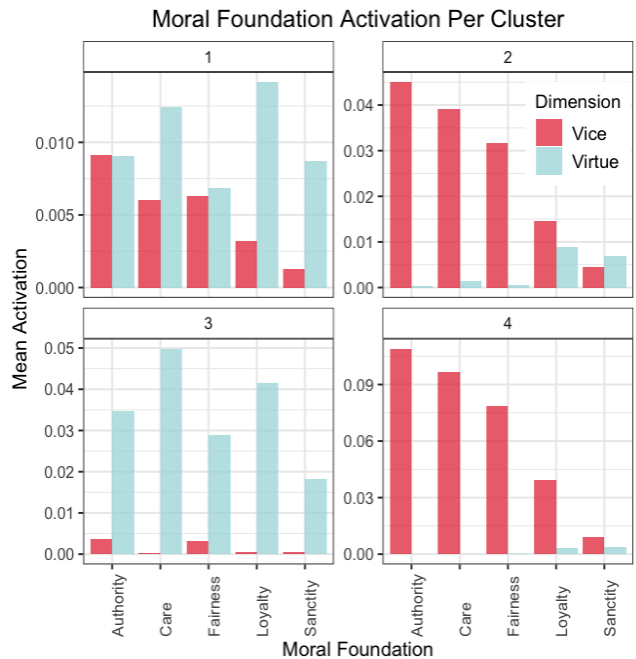

Figure 4. Mean activation for each *k*-means cluster of documents in embedding space.

| Cluster | Size | Cluster Label | Cluster Specific Vocabulary |
|---|---|---|---|
| 1 (Red) | 12,610 | Community Organizing | Los Angeles, California Hollywood LA |
| 2 (Teal) | 7,662 | Community Organizing | Los Angeles, California peaceful, protestors, protest |
| 3 (Blue) | 8,653 | Support for Movement | Support BLM, support, think, good, change |
| 4 (Black) | 1,298 | Hashtags | blacklivesmatter, georgefloyd, blackouttuesday |

Table 4. Mean activation for *k*-means each cluster of documents in embedding space.

## Discussion

We explored the moral sentiment of Black Lives Matter Tweets in Los Angeles after the killing of George Floyd. Quantitative methods mapped Tweets to a moral embedding space, which revealed that the most prominent moral concerns centered around discussions of (1) police authority and harm and (2) positive ingroup messaging. This work raises many questions about the role of moral expressions in light of extreme events, and how this can be used to better understand the offline reactions of a population. By more deeply understanding the moral framework of a population, we can better anticipate how they will respond to certain events. This will allow policymakers to better predict the outcome of their interventions.

These analyses have at least three limitations. First, the results are constrained to Tweets about one event geo-located to one city. Identical methods should be applied to discussions from other regions of the US about different, albeit similar, events. Second, more rigorous analyses of document clusters in the moral embedding space are due. As the present analysis was largely qualitative and exploratory, future work should develop quantitative metrics for examining clusters in the embedding space.

## Acknowledgements

Preparation for this manuscript was supported by ARO MURI grant W911NF1810208, NSF grant ATD-2027277, and the Defense Advanced Research Projects Agency (DARPA) and Army Research Office (ARO) under Contract No. W911NF-21-C-0002.